\documentclass[12pt,ams,a4paper,nofootinbib]{revtex4}
\usepackage{amsmath}
\usepackage{graphicx}
\usepackage[usenames,dvipsnames]{color}

\newcommand{\re}{\ref}
\newcommand{\be}{\begin{equation}}
\newcommand{\ee}{\end{equation}}

\newcommand{\la}{\label}
\newcommand{\ber}{\begin{eqnarray}}
\newcommand{\eer}{\end{eqnarray}}

\begin{document}

\title{Computing the coefficients of transformations between oscillator states}

\author{  Victor D. Efros$^{a,b}$
  }

\affiliation{
$^{a}$National Research Centre "Kurchatov Institute", 123182 Moscow, Russia\\
$^{b}$National Research Nuclear University MEPhI, 115409 Moscow, Russia 
}

\begin{abstract}
\noindent
{\bf Abstract.} 
A program is created to compute recursively the Moshinsky brackets. It is very fast
and provides highly accurate results. In the case of the double precision computations
with a single-processor consumer notebook,
the computing time per bracket at any not small oscillator excitations
is on the scale of 10$^{-8}$ s and the accuracy is very good for the total
number of
quanta up to 80. The program is easy to handle.

\end{abstract}

\bigskip

\maketitle

\section{Introduction}

The Talmi-Smirnov transformation of oscillator states
is widely applied in calculations 
in various branches of physics. The transformation is especially useful 
in the microscopic studies of structure of matter on various levels. An important 
nuclear physics application
is the calculation of matrix elements of nucleon-nucleon
 interaction in the oscillator shell-model basis,
see, e.g.,~\cite{mosh}.  

The coefficients  of the Talmi-Smirnov
transformation are called oscillator brackets.
At present,  large sets of oscillator states are frequently retained
in calculations. 
Therefore, 
a  fast program to compute oscillator  brackets is needed. The program should
provide accurate results 
up to high oscillator excitations.
In the present
work, such a program is presented. 

\section{preliminaries and definitions}

Let us write the Hamiltonian of the three-dimensional isotropic oscillator as 
\mbox{$\hbar\omega(q^2+x^2)/2$}
where ${\bf q}$ and ${\bf x}$ are canonically conjugated variables, 
\mbox{$[q_j,x_k]=-i\delta_{jk}$}.
Denote the normalized oscillator eigenstates as $|nlm\rangle$ where
$l$ and $m$  are the angular momentum and its projection, and $n$ is defined via
the eigenenergy \mbox{$\hbar\omega(2n+l+3/2)$}. 
The phase factors of the $|nlm\rangle$ states are chosen such that  
the ${\bf x}$-representation  wave functions are
\ber\langle {\bf x}|nlm\rangle=\left[\frac{2n!}{\Gamma(n+l+3/2)}\right]^{1/2}
x^{l}
L_n^{l+1/2}(x^2)
e^{-x^2/2}Y_{lm}({\bf n}_x)\la{cr}\eer
where the Laguerre polynomials $L_n^\alpha$ are defined in the standard way~\cite{gr},
and the usual 
spherical harmonics $Y_{lm}$ are employed
which form the standard basis with respect to rotations. 

Let  \mbox{$|n_1l_1n_2l_2LM_L\rangle$} be the states 
that are obtained via coupling the above oscillator states, 
belonging to different spaces, to a given
total angular momentum~$L$ and its projection~$M_L$,
\be|n_1l_1n_2l_2LM_L\rangle\equiv \sum_{m_1+m_2=M_L}C_{l_1m_1l_2m_2}^{LM_L}
|n_1l_1m_1\rangle|n_2l_2m_2\rangle.\la{co}\ee
They are eigenstates of the Hamiltonian
\be H=\hbar\omega(q_1^2+q_2^2+x_1^2+x_2^2)/2\la{H}\ee
with the eigenvalues \mbox{$\hbar\omega(N_q+3)$} where $N_q$ denotes
the total number of oscillator quanta, 
\be N_q=l_1+l_2+2(n_1+n_2).\ee
This notation is used below.

Let us perform a pseudo-orthogonal transformation
\ber {\bf x}_1={\bf x}_1'\cos\varphi+{\bf x}_2'\sin\varphi,\qquad
{\bf x}_2={\bf x}_1'\sin\varphi-{\bf x}_2'\cos\varphi.\la{pot}\eer
Denote the quantities canonically conjugated to ${\bf x}_1'$ and 
${\bf x}_2'$ as, respectively, ${\bf q}_1'$ and ${\bf q}_2'$. The relations 
expressing ${\bf q}_1$ and $\bf q_2$ in terms of ${\bf q}_1'$ and ${\bf q}_2'$
are of the same form as  Eqs.~(\re{pot}).
The Hamiltonian (\re{H}) preserves its form,
\be H=\hbar\omega[(q_1')^2+(q_2')^2+(x_1')^2+(x_2')^2]/2.\ee

Let $|n_1'l_1'm_1'\rangle$ and $|n_2'l_2'm_2'\rangle$ be, respectively, the eigenstates 
of the Hamiltonians  \mbox{$\hbar\omega[(q_1')^2+(x_1')^2]/2$}
 and $\hbar\omega[(q_2')^2+(x_2')^2]/2$. 
Let \mbox{$|n_1'l_1'n_2'l_2'LM_L\rangle$} be the eigenstates  of the Hamiltonian $H$ which are
constructed from the $|n_1'l_1'm_1'\rangle$ and $|n_2'l_2'm_2'\rangle$ states
in the same way as in Eq.~(\re{co}). 
The eigenstate \mbox{$|n_1l_1n_2l_2LM_L\rangle$} of Eq.~(\re{co})
may be expanded over  the 
\mbox{$|n_1'l_1'n_2'l_2'LM_L\rangle$}
eigenstates. 
Only those expansion states that have the same
energy as the original two-body harmonic oscillator state
enter the
expansion. 
Thus we have
\be |n_1l_1n_2l_2LM_L\rangle=\sum_{n_1'l_1'n_2'l_2'}
\langle n_1'l_1'n_2'l_2'|n_1l_1n_2l_2\rangle_L^\varphi\,|n_1'l_1'n_2'l_2'LM_L\rangle
\la{be}\ee
where the summation proceeds at the condition
\be l_1'+l_2'+2(n_1'+n_2')=l_1+l_2+2(n_1+n_2)=N_q.\la{ec}\ee
The expansion coefficients 
\be \langle n_1'l_1'n_2'l_2'|n_1l_1n_2l_2\rangle_L^\varphi\la{br}\ee
are called oscillator brackets or Talmi-Moshinsky-Smirnov coefficients (or brackets). 
Their computation
is addressed here. The present work is based on the results of 
Ref.~\cite{efr73} but the notation of that paper is modified here for convenience of
writing the computer 
program. The quantities $n_1'$ and $n_2'$ were denoted in Ref.~\cite{efr73} as
$N$ and $n$, the quantities $l_1'$ and $l_2'$ as $L$ and $l$, and the quantities $L$ and $M_L$
as $\lambda$ and $\mu$.

The transformation (\re{be}) is used, in particular, to calculate  
matrix elements of two-body
operators between the products of single-particle oscillator states.
If ${\bf r}_{1,2}$ are particle positions and ${\bf p}_{1,2}$ 
are their momenta then 
the above ${\bf x}_1$, ${\bf q}_1$, ${\bf x}_2$, and ${\bf q}_2$ quantities are
\ber {\bf x}_{1,2}={\bf r}_{1,2}\left[\hbar/(m_{1,2}
\omega)\right]^{-1/2},\qquad
{\bf q}_{1,2}={\bf p}_{1,2}(\hbar m_{1,2}\omega)^{-1/2}\la{trr}\eer
where $m_1$ and $m_2$ are masses of the particles. 

The above ${\bf x}_1'$, ${\bf q}_1'$, ${\bf x}_2'$, and ${\bf q}_2'$ quantities may be
taken as follows,
\ber {\bf x}_1'=({\bf r}_{2}-{\bf r}_{1})\left[\hbar/
(\mu\omega)\right]^{-1/2},\qquad{\bf q}_{1}'=
-{\bf p}_{rel}(\hbar \mu\omega)^{-1/2},\nonumber\\
{\bf x}_2'={\bf R}\left[\hbar/(M_0\omega)\right]^{-1/2},\qquad
{\bf q}_2'={\bf P}(\hbar M_0\omega)^{-1/2}.\la{trr1}\eer
Another choice is 
\ber
{\bf x}_1'={\bf R}\left[\hbar/(M_0\omega)\right]^{-1/2},\qquad
{\bf q}_1'={\bf P}(\hbar M_0\omega)^{-1/2},\nonumber\\
{\bf x}_2'=({\bf r}_{1}-{\bf r}_{2})\left[\hbar/
(\mu\omega)\right]^{-1/2},\qquad{\bf q}_{2}'=
{\bf p}_{rel}(\hbar \mu\omega)^{-1/2}.\la{trr2}\eer
In the above relations  $M_0=m_1+m_2$, $\mu=m_1m_2/M_0$ is the reduced mass, 
\mbox{${\bf p}_{rel}=(m_2{\bf p}_1-m_1{\bf p}_2)/M_0$} is the relative momentum,
\mbox{${\bf R}=(m_1{\bf r}_1+m_2{\bf r}_2)/M_0$}, and 
\mbox{${\bf P}={\bf p}_1+{\bf p}_2$}. These 
relations are the versions of the Smirnov~\cite{smi}
transformation. 

The above quantities are defined in a way that the
relations between (${\bf x}_1$, ${\bf x}_2$) 
and \mbox{(${\bf x}_1'$, ${\bf x}_2'$)} or \mbox{(${\bf q}_1$, ${\bf q}_2$)} 
and \mbox{(${\bf q}_1'$, ${\bf q}_2'$)} are, indeed, of the form of Eqs.~(\re{pot}). 
In the case of Eqs.~(\re{trr1}) one gets
\be \cos\varphi=-\left(\frac{m_2}{m_1+m_2}\right)^{1/2},\qquad
\sin\varphi=\left(\frac{m_1}{m_1+m_2}\right)^{1/2}.\la{m}\ee
In the case of Eqs.~(\re{trr2}) one gets
\be \cos\varphi=\left(\frac{m_1}{m_1+m_2}\right)^{1/2},\qquad
\sin\varphi=\left(\frac{m_2}{m_1+m_2}\right)^{1/2}.\la{m1}\ee 
(This is seen at once if the inverse transformations are considered.) 

The expansions (\re{be}) are also of use  when solving three-particle problems in the
oscillator basis. In the case of identical particles these expansions
 make possible to construct basis states
that behave properly under particle permutations. 
If  not all three particles are identical these expansions are helpful 
to calculate interaction
matrix elements. This is discussed, e.g., in Ref.~\cite{efr20} as to 
the hyperspherical case.

If instead of the transformation (\re{pot}) the orthogonal transformation
\ber {\bf x}_1={\bf x}_1'\cos\varphi+{\bf x}_2'\sin\varphi,\qquad
{\bf x}_2=-{\bf x}_1'\sin\varphi+{\bf x}_2'\cos\varphi\la{opot}\eer
is considered  then the coefficients of
the corresponding expansion 
similar to (\re{be}) are
obviously expressed in terms of the coefficients (\re{br}) as
\be(-1)^{l_2}\langle n_1'l_1'n_2'l_2'|n_1l_1n_2l_2\rangle_L^\varphi.\la{otr}\ee
Use of the transformation (\re{pot}) is preferable  since in this case
a simple symmetry relation 
\be \langle n_1'l_1'n_2'l_2'|n_1l_1n_2l_2\rangle_L^\varphi=
\langle n_1l_1n_2l_2|n_1'l_1'n_2'l_2'\rangle_L^\varphi \la{sym}\ee
is valid. This has been pointed out in Ref.~\cite{bd} 
for the case of equal masses ($m_1=m_2$)  in Eqs.~(\re{trr}) and (\re{trr2})
and in Ref.~\cite{efr73} in the general form (\re{sym}). As mentioned in~\cite{efr73} 
the relation (\re{sym})
follows
from general properties of the transformation (\re{be}) which is
 commented also in~\cite{efr20}.
We shall deal with the transformation~(\re{pot}) and not with~(\re{opot})
in what follows.

Other symmetry relations for the brackets also exist~\cite{efr73} 
(as to the  $m_1=m_2$  case see also~\cite{bd,mosh1}). 
The symmetry relations for oscillator 
brackets presented  as being original in~\cite{gk} are in fact
the same as those given previously in~\cite{efr73}.

\section{Relations to calculate the brackets}

Moshinsky~\cite{mosh}
has suggested the following algorithm to calculate oscillator brackets. First the brackets 
with \mbox{$n_1=n_2=0$} are calculated. Next   
the rest brackets are obtained
with the help of \mbox{$n_1-1\rightarrow n_1$} and \mbox{$n_2-1\rightarrow n_2$}
recurrence relations. He obtained such  recurrence relations and an expression for the
\mbox{$n_1=n_2=0$} brackets. 
He provided the formulae in the case of  the transformation 
(\re{opot}) at \mbox{$\varphi=\pi/4$} (which corresponds to the $m_1=m_2$ case in
Eq.~(\re{trr1}) type relations).  
Below we shall refer to such formulae 
pertaining to the transformations (\re{pot}) and (\re{opot}) at any $\varphi$
 as to
Moshinsky-type relations.

In Ref.~\cite{efr73} the recurrence relations realizing the Moshinsky algorithm have
 been obtained
in a form different from the Moshinsky-type form. Also an expression for the
initial \mbox{$n_1=n_2=0$} brackets different from the Moshinsky-type expression
has been derived 
there.
The
present program is based on the
corresponding relations of Ref.~\cite{efr73}, Eqs.~(\re{rr}) and (\re{ibr}) of the present paper.

The recurrence formulae in~\cite{efr73} have been obtained for modified brackets, 
 denoted as 
$[n_1'l_1'n_2'l_2'|n_1l_1n_2l_2]_L^\varphi$ and defined as follows, 
\be  [n_1'l_1'n_2'l_2'|n_1l_1n_2l_2]_L^\varphi= 
\langle n_1'l_1'n_2'l_2'|n_1l_1n_2l_2\rangle_L^\varphi\frac{A(n_1',l_1')A(n_2',l_2')}
{A(n_1,l_1)A(n_2,l_2)}\la{mbr}\ee
where 
\be A(n,l)=(-1)^n[(2n)!!(2n+2l+1)!!]^{-1/2}.\la{anl}\ee
Such a modification of brackets was done in~\cite{bd} in a different context and 
in~\cite{se} in case of the hyperspherical brackets. 
 
The $n_1-1\rightarrow n_1$ recurrence formula is
\ber [n_1'l_1'n_2'l_2'|n_1l_1n_2l_2]_L^\varphi=\cos^2\varphi
[n_1'-1l_1'n_2'l_2'|n_1-1l_1n_2l_2]_L^\varphi\nonumber\\+\sin^2\varphi
[n_1'l_1'n_2'-1l_2'|n_1-1l_1n_2l_2]_L^\varphi\nonumber\\
-\sin\varphi\cos\varphi\biggl
\{[n_1'l_1'-1n_2'l_2'-1|n_1-1l_1n_2l_2]_L^\varphi\,\psi_L^+(l_1',l_2')\nonumber\\
+[n_1'-1l_1'+1n_2'-1l_2'+1|n_1-1l_1n_2l_2]_L^\varphi\,\psi_L^+(l_1'+1,l_2'+1)\nonumber\\
-[n_1'-1l_1'+1n_2'l_2'-1|n_1-1l_1n_2l_2]_L^\varphi\,\psi_L^-(l_1'+1,l_2')\nonumber\\
-[n_1'l_1'-1n_2'-1l_2'+1|n_1-1l_1n_2l_2]_L^\varphi\,\psi_L^-(l_1',l_2'+1)\biggr\}\la{rr}\eer
where
\be \psi_L^\pm(p,q)=\left[\frac{[(p\pm q)^2-L^2][(p\pm q)^2-(L+1)^2]}{(4p^2-1)(4q^2-1)}
\right]^{1/2}.\la{psi}\ee
(In Ref.~\cite{efr73} the quantities $\psi_L^+(p,q)$ were
 denoted as $\alpha_\lambda(p,q)$ and  
$\psi_L^-(p+1,q)$ as $\beta_\lambda(p,q)$. One then has 
\mbox{$\psi_L^-(p,q+1)=\psi_L^-(q+1,p)=\beta_\lambda(q,p)$}).

The \mbox{$n_2-1\rightarrow n_2$} recurrence formula 
is obtained from Eq.~(\re{rr}) via making
the replacements \mbox{$\cos^2\varphi\leftrightarrow\sin^2\varphi$} and 
changing the sign of $\sin\varphi\cos\varphi$.

The \mbox{$[n_1'l_1'n_2'l_2'|0l_10l_2]_L^\varphi$} coefficients are calculated
separately and they make possible
to start the recursion. They are given by the following expression,\footnote{Eq.~(\re{ibr})
is Eq.~(39)  in~\cite{efr73} but it is written here for the modified coefficient 
\mbox{$[\ldots|0l_10l_2]_L^\varphi$}
instead of \mbox{$\langle \ldots|0l_10l_2\rangle_L^\varphi$}. Let us mention
that the most simple derivation of the corresponding Eq.~(23) in~\cite{efr73} is contained
in Appendix there.}
\ber
[n_1'l_1'n_2'l_2'|0l_10l_2]_L^\varphi=(-1)^{n_1'}
\prod_{i=1}^22^{-l_i'}\left[(2l_i+1)(2l_i'+1)\right]^{1/2}\left[A(n_i',l_i')\right]^2
\nonumber\\
\times
[(l_1+l_2+L+1)!(l_1+l_2-L)!]^{1/2}(\cos\varphi)^{l_1+l_2}(\tan\varphi)^{n_1'+n_2'}
F_L^\varphi
\la{ibr}\eer
where $A(n_i',l_i')$ are from Eq.~(\re{anl}), and
\be
F_L^\varphi=\sum_{i=i_{min}}^{i_{max}}(-1)^{\alpha_{i4}}
\frac{\left[\prod_{k=1}^4(2\alpha_{ik})!\right]^{1/2}}{\prod_{k=1}^4\alpha_{ik}!}
\left(\begin{array}{ccc}
{\tilde l}_1&{\tilde l}_2&L\\
l_1'-i&i-l_2'&l_2'-l_1'
\end{array}\right)\tan^i\varphi.\la{fl}\ee
Here $(\ldots)$ is the $3j$-symbol,
\[ {\tilde l}_1=\frac{l_1'+l_2'+l_1-l_2}{2},\qquad {\tilde l}_2=\frac{l_1'+l_2'-l_1+l_2}{2},\]
\be \alpha_{i1}=\frac{{\tilde l}_1+l_1'-i}{2},
\quad\alpha_{i2}=\frac{{\tilde l}_1-l_1'+i}{2},\quad\alpha_{i3}=
\frac{{\tilde l}_2-l_2'+i}{2},\quad
\alpha_{i4}=\frac{{\tilde l}_2+l_2'-i}{2}.\la{alp}
\ee
 The summation proceeds between the limits
\[ i_{min}=|{\tilde l}_1-l_1'|\equiv|{\tilde l}_2-l_2'|,\qquad i_{max}=
\min({\tilde l}_1+l_1',{\tilde l}_2+l_2')\]
within which the $3j$-symbol is different from zero. The summation variable takes only
values of the same parity as these limits. This is related to the requirement that the
quantities~(\re{alp}) must be integers. (These quantities are non-negative.)

A remarkable feature of the recurrence formula (\re{rr})
 is that at a given $L$ value the coefficients $\psi_L^\pm$
depend  on only two quantum numbers $l_1'$ and $l_2'$ and
are independent of the quantum
numbers $n_1'$, $n_2'$, $n_1$, $n_2$, $l_1$, and $l_2$. In addition, 
these coefficients are
expressed in terms of only two 
functions of Eq.~(\re{psi}). 
All this suggests precomputing  these two functions which is very fast.
Different from the above
 recurrence formulae, the Moshinsky-type recurrence formulae involve 
the initial brackets (\re{br}). The coefficients of such formulae are 
products of six different
functions depending on
 four quantum numbers and functions depending on two quantum numbers. 

The quantity (\re{fl}) depends on  $l_1'$, $l_2'$, and on the difference $l_1-l_2$.
In our case,
precomputing this quantity also proves to be helpful. 
The Moshinsky-type derivation leads to a different formula for the \mbox{$n_1=n_2=0$}
 brackets 
which 
includes full dependence on the four partial 
angular momenta and at  given their values involves probably two-three times more 
computations than in case of Eq.~(22).

\section{angular momentum variables}

Permissible values of the  quantum numbers $l_1$ and~$l_2$, or $l_1'$ and~$l_2'$,
specifying a bracket
 are restricted by the    requirement of a given parity and
 by the triangle inequalities involving the $L$ value.
 Aiming both to  get compact arrays of brackets  and avoid the 
 restrictions in computations  
 we shall use the following variables instead of $l_1$ and~$l_2$, or
 $l_1'$ and~$l_2'$,
 \ber M=\frac{l_1+l_2-L-\epsilon}{2},\quad N=\frac{l_1-l_2+L-\epsilon}{2},\nonumber\\
 M'=\frac{l_1'+l_2'-L-\epsilon}{2},\quad N'=\frac{l_1'-l_2'+L-\epsilon}{2}\la{var}
 \eer
 where $\epsilon=0$ or 1 when $N_q-L$ is even or odd, respectively. The inverse relations are
 $l_1=M+N+\epsilon$, $l_2=M-N+L$ and $l_1'=M'+N'+\epsilon$, $l_2'=M'-N'+L$.
 These variables have been introduced in~\cite{efr21} in the case of hyperspherical brackets.
 One has 
  \be (N_q-L-\epsilon)/2=M+n_1+n_2=M'+n_1'+n_2'.\la{nml}\ee
  
  When $l_1$, $l_2$, $l_1'$, and $l_2'$ take all the values
  allowed at a given $L$ and a given parity, the $M$ or~$M'$ and $N$ or~$N'$ 
  variables take all the integer
  values from zero up to, respectively, 
  \be M_{max}=(N_q-L-\epsilon)/2,\quad N_{max}=L-\epsilon.\la{max}\ee
  (The case $L=0$, $\epsilon=1$ is, obviously, not possible.) Thus the discussed variables
  densely fill intervals independent of each other.

\section{The programs}

We suggest that often the most convenient form of the output 
 would be an array, or 'table', of
 all the brackets pertaining to states 
 with $N_q$ values up to some $(N_q)_{max}$, with
total angular momenta in a  range from $L_{min}$ to $L_{max}$,
and with a given parity.
In particular, if
such an array  is available then one can avoid repeated computations of the same
brackets at the calculation of matrix elements. In the framework of
 the present algorithm, such an
 array is composed of groups of brackets having the same
$l_1$, $l_2$, and $L$ values. The brackets of each group are related by the recursion.

In the present work, brackets sought for are produced by a subroutine written in two versions.
 One of them is   named  \mbox{ALLOSBRAC} and its output is just the mentioned array.

The parameters of the \mbox{ALLOSBRAC} subroutine  
are NQMAX, LMIN, LMAX, CO, SI, and BRAC. Here
NQMAX, LMIN, and LMAX  
are the input parameters
defined above. (NQMAX $\equiv (N_q)_{max}$, LMIN $\equiv L_{min}$, etc. 
We use capital letters for quantities entering
the programs.) 
CO and SI are the input parameters $\cos\varphi$ and 
$\sin\varphi$ from Eq.~(\re{pot}). BRAC is the output array of brackets.
 Values of $N_q$ pertaining to all these brackets are such that
$(N_q)_{max}-N_q$ is even.

In the present programs, the quantities like~$N'$, $n_1'$, etc., are denoted like NP, 
N1P, etc., where P
symbolizes 'primed'. 
The mentioned output  array BRAC is of the form \mbox{BRAC(NP,N1P,MP,N1,N2,N,M,L)}.
 The value of N2P is determined
by Eq.~(\re{nml}). Thus the arguments of BRAC
represent all quantum numbers specifying a bracket. The order of arguments of BRAC 
corresponds to the nesting of loops in the subroutine
 at the 
 computation of brackets 
 since such an order is preferable. In accordance with the bounds 
 of the BRAC array in the dimension list,
 the size of this array is
\[(L_{max}-L_{min}+1)(L_{max}+1)^2\{{\rm Int}[[(N_q)_{max}-L_{min}]/2]+1\}^5.\]
In many cases such a size is acceptable. (At the double precision calculations
when \mbox{$(N_q)_{max}=40$} and  
\mbox{$L_{min}=L_{max}=L$} this size is about or less 1 GB at any $L$. When 
\mbox{$(N_q)_{max}=L_{max}=20$}
and \mbox{$L_{min}=0$} it is about 12 GB.)

The other version of the subroutine, named 
 \mbox{OSBRAC}, produces
 the above mentioned subsets of brackets with given
 $l_1$, $l_2$, and $L$ values or, which is the same, with given $M$, $N$, and $L$ values.
 As mentioned above, these brackets pertain to states 
 with $N_q$ values up to some $(N_q)_{max}$
and with given parity determined by the $(N_q)_{max}$ value.
This version 
 is required, in particular, in the cases of memory restrictions. 
 (Of course,  one can also compute an array at given 
 $l'_1$ and $l_2'$ values instead of given $l_1$ and $l_2$ values, see (\re{sym}).)
 
The parameters of the   \mbox{OSBRAC} subroutine   
are N, M, L, NQMAX, CO, SI, \mbox{FIRSTCALL}, and BRAC. 
All the listed parameters but BRAC are input ones. 
All of them  except \mbox{FIRSTCALL} represent the quantities defined above. The
\mbox{FIRSTCALL} parameter is discussed below.
 The BRAC parameter represents the output array of brackets. 
In this case, it is of the form BRAC(NP,N1P,MP,N1,N2). It thus represents the subset
of brackets pertaining to given input N, M, and L values. Again, N2P is determined
by Eq.~(\re{nml}) and the arguments of BRAC along with N, M, and L represent all 
quantum numbers specifying a bracket. 

Describing the 
 programs, let us first discuss  the issue of possible overflows
at large quantum numbers. When the double precision is set in the present
programs, the allowed
values of NQMAX are as follows, $(N_q)_{max}\le84$. 
At this restriction, the computation as real numbers of
 all the factorials and double factorials entering Eqs.~(\re{anl}), (\re{ibr}),
and (\re{fl}), apart from
the 3$j$ symbols in Eq.~(\re{ibr}),  does not lead to overflows/underflows. 
(To conclude this, one needs to take
 into account that in Eq.~(\re{alp}) $\alpha_{ik}\le N_q$.)
This restriction may be considerably weakened if square roots of factorials are calculated
with the quadrupole precision
and all the rest with the double precision.  This is not done here 
because, 
at $(N_q)_{max}$ values already not much higher than 84,
 accuracy of the present double precision computation 
becomes  insufficient in general, 
see Table V in the next section. Let us mention in this
connection that e.g., passing  from the double precision to the quadrupole precision  everywhere
 in the calcultion of
Ref.~\cite{lit2} 
increased the running time  more than 15 times. 

Still, products of several
factorials entering the usual expressions for 3$j$ symbols  may lead to
overflows. To avoid this, we employed the expression 
\cite{jap} (see also \cite{var}) for 
3$j$ symbols in terms of binomial coefficients,
\ber \left(\begin{array}{ccc}
j_1&j_2&j_3\\
m_1&m_2&m_3
\end{array}\right)=(-1)^{j_1-j_2-m_3}\left(\frac{Bi(\gamma_2,2j_1)
Bi(\gamma_3,2j_2)}{(2j_3+1)Bi(\gamma_3,j_1+j_2+j_3+1)\Pi}\right)^{1/2}S,\nonumber\\
\Pi=Bi(j_1+m_1,2j_1)Bi(j_2+m_2,2j_2)Bi(j_3+m_3,2j_3)
,\nonumber\\
S=\sum_{k=k_{min}}^{k_{max}}(-1)^kBi(k,\gamma_3)
Bi(j_1-m_1-k,\gamma_2)Bi(j_2+m_2-k,\gamma_1)\la{jap}\eer
where $Bi(m,n)=n!/[m!(n-m)!]$ are the binomial coefficients, $\gamma_i=j_1+j_2+j_3-2j_i$, and
\[k_{min}={\rm max}(0,j_1-j_3+m_2,j_2-j_3-m_1),\quad
k_{max}={\rm min}(\gamma_3,j_1-m_1,j_2+m_2).\]
We note that if one writes the binomial coefficients entering here in terms of factorials then
Eq.~(\re{jap}) turns to the standard Van der Waerden formula~\cite{var} for 3$j$
symbols. In Ref.~\cite{kit} it has been realized that Eq.~(\re{jap}) type expressions 
are advantageous for
computations. In Ref.~\cite{wei}
their utility to cope with the overflow issue has been noted.  
Due to the fact that
 $Bi(m,n)$ grow with the increase of $n$ slower than factorials, $Bi(m,n)\le2^n$, 
no overflows can arise  in double precision computations
at use of the expression (\re{jap})
with the above restriction on $(N_q)_{max}$. 

The structures and details
of the present programs should be clear from generous comments
they include.
Below an outline is presented.
Both subroutines \mbox{ALLOSBRAC} and \mbox{OSBRAC}
first call for the simple subroutines  ARR and COE that precompute the required factorials, 
double factorials, quantities $[(2I)!]^{1/2}/I!$ from Eq.~(\re{fl}), and the binomial coefficients. 
Then the quantities~(\re{anl})  are precomputed.  
Next both \mbox{ALLOSBRAC} and \mbox{OSBRAC} call for the subroutine named
\mbox{FLPHI} to precompute
the array of the quantities~(\re{fl}) multiplied by $(2l_1'+1)(2l_2'+1)$. 
The \mbox{FLPHI} subroutine
calls for the function WIGMOD that provides the quantity $S$ from Eq.~(\re{jap})
times the factors from there depending on $m_1$ and $m_2$. 
Furthermore, to precompute the arrays 
representing the coefficients (\re{psi}) of the recurrence formulae,
the subroutine named \mbox{COEFREC} is called. In the \mbox{ALLOSBRAC} case,
\mbox{FLPHI} and \mbox{COEFREC} are called inside the loop over $L$.

All the mentioned routines are contained in the file~allosbrac.f90 in the case of
the first mentioned version of the program and in the file~osbrac.f90 in the case of
the second one. 

The recurrence procedures are realized as follows. First  the recursion
$n_2-1\rightarrow n_2$ discussed in Sec. III
is performed at $n_1=0$. Then the recursion $n_1-1\rightarrow n_1$
is performed at each $n_2$ value starting from the obtained brackets with $n_1=0$.

 One should require that the arguments
of the array BRAC in
the right-hand  sides of the recurrence relations like (\re{rr})
do not go beyond their bounds prescribed at the preceding step of the recursion. 
When this condition is not fulfilled, the corresponding contributions to the recurrence
relations are to be omitted. The arising restrictions are summarized 
in comment lines in the subroutines.  One case of the restrictions is as follows.

Consider a step of the recursion. According to Eq.~(\re{nml}), the N1P and MP
values pertaining to this step are such that
\mbox{N1P+MP $\le M_{max}$}. 
The $N_q$ value increases by two at each step of the recursion. Correspondingly,
the  $M_{max}$ value given by Eq.~(\re{max}) increases by one. 
Let us write the array BRAC as
BRAC(K1,K2,K3\ldots) where K2 and K3 
represent, respectively, the  N1P and MP
variables. 
Then, according to the said above, for
the array BRAC in the right-hand sides of the recurrence relations the condition
\mbox{K2+K3 $\le M_{max}-1$} should be fulfilled since 
the array BRAC there corresponds 
to the preceding step of the recursion. 
At given N1P and MP values mentioned above, possible K2 and K3 values pertaining to
the preceding step of the recursion are the following,
K2=N1P or \mbox{N1P-1} and K3=MP or MP-1, or MP+1. This is seen from Eq.~(\re{rr}) 
rewritten in terms of
the $M'$ and $N'$ variables.  

All the combinations of these K2 and K3 values except for (K2,K3)=(N1P,MP+1)
are present in the recurrence relations. 
When \mbox{(K2,K3)=(N1P,MP-1)}, or \mbox{(K2,K3)=(N1P-1,MP-1)}, 
or \mbox{(K2,K3)=(N1P-1,MP)}
the above condition \mbox{K2+K3 $\le M_{max}-1$} fulfills automatically.
But when \mbox{(K2,K3)=(N1P,MP)} or \mbox{(K2,K3)=(N1P-1,MP+1)} 
the requirement 
\mbox{N1P+MP $\ne M_{max}$} is to be imposed in order this  condition fulfills. (The  restriction we described matters if 
 the BRAC array includes non-zero elements 
 at the time of the call of the OSBRAC or ALLOSBRAC routine.) 

The
FIRSTCALL parameter of the \mbox{OSBRAC} subroutine
is a logical variable.  
 This variable makes possible not to call for the subroutine \mbox{FLPHI} by
 \mbox{OSBRAC} at all, except for the first, sequental calls for 
  \mbox{OSBRAC} with a given $L$ value and with each of $N$ (i.e., $l_1-l_2$) 
  values.
To achieve this,
it is to be said \mbox{FIRSTCALL=.FALSE.} 
in the proper place of the program that calls for \mbox{OSBRAC}.
Then, a change of $L$  
in that program is to be accompanied by the statement 
\mbox{FIRSTCALL=.TRUE.} Examples of use of the variable \mbox{FIRSTCALL} 
are given in the appended program \mbox{TESTOSBRAC}, tests 3 and 4.
(The statement \mbox{FIRSTCALL=.FALSE.} also  removes most of unnecessary calls
of other subroutines by \mbox{OSBRAC}   
   but this is not important.)

Of course, there exists also a possibility not to use the described option and to call for
\mbox{OSBRAC} always at  \mbox{FIRSTCALL=.TRUE.} The effect of the variation of
FIRSTCALL is not very large, see the last section.

\section{tests}

Tests can be readily performed 
with the help of the appended programs \mbox{TESTALLOSBRAC}
 and \mbox{TESTOSBRAC}.
 In addition to checks of brackets, 
these programs provide commented examples
of implementation of \mbox{ALLOSBRAC} and \mbox{OSBRAC}.
Outcomes of the tests
are contained, respectively,  in the files named \mbox{allosoutput} and \mbox{osoutput}.
The tests as well as the calculations in the next section were performed 
at \mbox{$m_1=m_2$} in Eq.~(\re{m}). The tests were the following. 

1. For the brackets with \mbox{$n_2'=l_2'=0$} and thus with \mbox{$l_1'=L$}
 and \mbox{$n_1'=(N_q-L)/2$}
there exists a simple explicit expression~\cite{efr19}\footnote{This expression has
been derived in~\cite{efr19} for the case of pseudo-orthogonal transformations~(\re{opot}).
For orthogonal transformations~(\re{pot}) considered here the result is the same and the
derivation is a bit simpler due to the symmetry relation  (\re{sym}).}
\ber \langle n_1'L00|
n_1l_1n_2l_2\rangle_L^\varphi=\cos^{2n_1+l_1}\varphi\sin^{2n_2+l_2}\varphi
\nonumber\\
\times
(-1)^L\left[(2l_1+1)(2l_2+1)\right]^{1/2}
\left(\begin{array}{ccc}l_1&l_2&L\\0&0&0\end{array}\right)
\frac{A(n_1,l_1)A(n_2,l_2)}{A(n_1',L)}\la{br0}\eer 
where the notation (\re{anl}) is used. The bracket computed 
with the programs has been compared with this expression.

2. Symmetry (\re{sym}) of brackets computed with the programs has been verified. 

3. Consider the relation 
\be\sum_{i,i',j,L;N_q=const}\langle j|i\rangle_L^\varphi
\langle j|i'\rangle_L^\varphi
=\nu(N_q).
\la{ns}\ee
Here $i$, $i'$, and  $j$ symbolize various     $n_1l_1n_2l_2$  sets.
 The summations proceed over all such sets existing at a given $N_q$ value. 
The quantities 
$\langle..|..\rangle_L^\varphi$ thus represent the brackets we deal with. 
The quantity $\nu(N_q)$ is the number of  all
 the \mbox{$|n_1l_1n_2l_2LM_L\rangle$} states, with the same  $M_L$,
 existing at a given $N_q$ value.

 This number is $\sum_{{\rm all}\,L}\nu(N_q,L)$ where $\nu(N_q,L)$ is  
 the number of all such states  at given $N_q$ and given 
$L$ values.  
The latter number equals \mbox{$(N_{max}+1)(M_{max}+1)(M_{max}+2)/2$}
where $N_{max}$ and $M_{max}$ are given by Eq.~(\re{max}). Indeed, 
according to
 Eq.~(\re{nml}) the number of such states at a given $M$ value
 with all possible $N$ values  
equals \mbox{$(N_{max}+1)[(M_{max}-M)+1]$}. 

Thus the left-hand 
side of Eq.~(\re{ns}) computed with the 
programs  has been  verified. 

4. The  quantity equal to zero
\be
\delta=\sum_{i,i',L;N_q\le(N_q)_{max}}\bigl|\bigl[\sum_j
\langle j|i\rangle_L^\varphi\langle j|i'\rangle_L^\varphi\bigr]
-\delta_{ii'}\bigr|\la{del}
\ee
has been computed with the programs.
Here the notation is as in Eq.~(\re{ns}) and
the summations proceed over all the brackets with $N_q$ values 
that do not exceed some $(N_q)_{max}$.

\section{accuracies and running times}

All the calculations below have been performed with a consumer notebook 
Intel core of the first generation \mbox{i5-750~2.67~GHz} (2009).
Since the computer had a small active memory (3.6~GB) 
the \mbox{ALLOSBRAC} routine was employed at $L_{min}=L_{max}=L$
to make the calculation executable in all the cases. 
 To determine small running times, the computations were
done repeatedly. Times to print out the outputs were disregarded. 
In the present computations and those of other authors discussed below  
the
double precision has been set.

Programs to calculate oscillator brackets have been published in 
\mbox{Refs.~\cite{gk,lit2,lej,sg,zo,ft,do,lit3}}.
Of them, the programs~\cite{gk,lit2,lit3} are written in a contemporary programming language.
They are based on an explicit expression~\cite{buck} for the brackets. 
Comparison of the results of the present  calculations with those of Refs.~\cite{gk,lit2,lit3} 
is done below.

The programs of Refs.~\cite{lej,sg,zo,ft,do} are written in Fortran - IV (1965). 
The too complicated program~\cite{lej} is based on the formulae of Ref.~\cite{mosh} 
(taken from~\cite{bro}).
It is applicable only in a limited range of quantum numbers. 
It was stated~\cite{ft}, see also~\cite{bd},
 that the recurrence relations of Ref.~\cite{mosh} 
 are very time consuming and somewhat inefficient to compute
the brackets. 
In the programs of \mbox{Refs.~\cite{sg,zo,ft,do}} 
explicit expressions for brackets~\cite{smi,bd,trl,do1} have been 
employed.\footnote{Methods to calculate the brackets were also worked out  in 
Refs.~\cite{ar,smi1,kum,bak,tal,ray,do2}. Refs.~\cite{smi1,kum,bak}
deal with explicit formulae, Ref.~\cite{tal} deals with the matrix diagonalization, and
Refs.~\cite{ar,ray,do2} deal with the recurrence relations different from the present ones.}
In~\cite{ft,do} ingredients  of such expressions which are
independent of $n$'s
were precomputed. This speeds up the computations 
in the cases when sets of brackets with many $n$'s and limited $l$'s are required.
In the case of the programs of Refs.~\cite{lej,sg,zo,ft,do} no information 
on stability  of the computation algorithms  is available. Contrary to our results below,
only brackets at rather low oscillator excitations were
considered in the mentioned papers.

\begin{table}[ht]
\caption{Accuracy and running times at calculating the quantity $\delta$ of Eq.~(\re{del})
in comparison with Ref.~\cite{gk}. First column: $(N_q)_{max}$ value in Eq.~(\re{del}). 
Second column: the 
values of $\delta$ obtained. 
Third column: running times~$t_1$ in seconds calculated 
with \mbox{ALLOSBRAC}. Fourth column:
running times~$t_2$ in seconds calculated with \mbox{OSBRAC}. Fifth column: the values of
Ref.~\cite{gk} for $\delta$ denoted as $\delta'$. Last column: the ratios
of the running time $t'$ of Ref.~\cite{gk} to that in the third column. These
ratios are commented  in 
the text.}
\begin{tabular}{|c|c|c|c|c|c|}
\hline
$(N_q)_{max}$&$\delta$&$t_1$, $[s]$
&
$t_2$, $[s]$
&$\delta'$~\cite{gk}&
$t'/t_1$
\\
\hline
8&$7.9\cdot10^{-13}$&$1.9\cdot10^{-4}$&$1.3\cdot10^{-3}$
&$7.0\cdot10^{-10}$&$9.2\cdot10^{6}$\\
\hline
12&$6.8\cdot10^{-12}$&$1.3\cdot10^{-3}$&$2.0\cdot10^{-2}$
&$3.5\cdot10^{-5}$&$1.9\cdot10^{8}$\\
\hline
16&$3.5\cdot10^{-11}$&$6.8\cdot10^{-3}$&0.19&-&-\\
\hline
20&$1.4\cdot10^{-10}$&$2.7\cdot10^{-2}$&1.2&-&-\\
\hline
24&$4.75\cdot10^{-10}$&$9.1\cdot10^{-2}$&5.65&-&-\\
\hline
28&$1.5\cdot10^{-9}$&0.28&22&-&-\\
\hline
\end{tabular}
\end{table}

In Table I the computed values of the sum $\delta$ of Eq.~(\re{del}) and the corresponding
running times
are presented for a number
of $(N_q)_{max}$ values. The computations were performed with both \mbox{ALLOSBRAC}
 and \mbox{OSBRAC}. The calling programs here are the same  as   
 in  the  appended files mentioned in the preceding section. The contributions of even and odd
 $N_q$ values were added up.
  (In the \mbox{OSBRAC} case arrays 
 of brackets  were not stored. Otherwise, the calculation would had been
  equivalent to use of \mbox{ALLOSBRAC}.) The $\delta$ values
produced by the two programs coincide with each other. This should be the case 
since eventually  the same operations 
are performed in the same order in the two computations. (Although in the \mbox{OSBRAC}
case some repetitions occur.)
The $\delta$ values obtained are shown in the  second column. 
The running times in seconds  pertaining to \mbox{ALLOSBRAC}
 and \mbox{OSBRAC} are presented in the third and fourth column, respectively.
They are denoted as $t_1$
and~$t_2$.  It is seen that at higher $(N_q)_{max}$ values \mbox{OSBRAC}
is much slower. This is because of the repeated calculations
of the same brackets required in the \mbox{OSBRAC} case. If the varying 
of FIRSTCALL is not applied
then the running time~$t_2$ becomes more than twice larger than
in the table at $(N_q)_{max}=12$ and by 30\% larger at $(N_q)_{max}=28$.

The values of $\delta$ obtained    in Ref.~\cite{gk}
are listed in the fifth column and denoted as $\delta'$. 
It is seen that the algorithm based on an explicit expression for the brackets
employed  in Ref.~\cite{gk} leads to
much faster deterioration of accuracy
with the increase of $(N_q)_{max}$ than in the case of the present approach.

In the last column the ratios 
of the running times of Ref.~\cite{gk}, denoted as $t'$, to
the present running times $t_1$ are listed. 
Discussing them, we shall take into account that, as it can be seen,  
the contribution    to the considered
running times of summations in Eq.~(\re{del}) 
can be disregarded.
The drastic difference in the running times is caused by several reasons and one 
of them is the following. 
 The numbers of 
 computed brackets listed in Ref.~\cite{gk} show that the same brackets
 were computed there repeatedly for many times.
And in case of use of
our \mbox{ALLOSBRAC} routine all the brackets are computed only once.
 But this is not the only point, and
also average running times per one computed bracket are 
much larger in the case of Ref.~\cite{gk} than in the present case. 
The ratio of these  running times equals
 $8.2\cdot10^3$ 
at $(N_q)_{max}=8$
and $4.5\cdot10^4$  at $(N_q)_{max}=12$. Probably, in accordance with the Moore's law,  
about two digits
in these ratios may 
be attributed to the difference between the performances of the computer used 
in the present calculation
and of 
the older PC of Ref.~\cite{gk}. If one accepts this, then it may be concluded 
that the average number of operations to calculate a bracket in
Ref.~\cite{gk}  is  larger  more than 80 times at  
\mbox{$(N_q)_{max}=8$} and  about 450 times at
\mbox{$(N_q)_{max}=12$} than in 
the case of the present program.

In Table II the running times $t_{\scriptscriptstyle\Sigma}$ 
are listed which pertain to the calculation of all existing brackets  with
a given $L$ value and with $N_q$ values such that $N_q\le (N_q)_{max}$ and
$(N_q)_{max}-N_q$ is even. 
These running times refer to use of \mbox{ALLOSBRAC}.
When \mbox{OSBRAC} is used the running times are almost the same. This is 
because the number of
arithmetic operations performed is the same  in this case
except for some  repetitons of fast precomputations.
See also the next table in this connection.

\begin{table}[ht]
\caption{Third column: the running times $t_{\scriptscriptstyle\Sigma}$ 
in seconds for the calculation of
 the sets of all brackets with a given $L$ value 
 and with $N_q$ values such that $N_q\le (N_q)_{max}$ and
$(N_q)_{max}-N_q$ is even. 
Last column: the ratios  $t'/t_{\scriptscriptstyle\Sigma}$,
where $t'$ are the running
times  of Ref.~\cite{lit2}, for the calculation of the brackets with 
 $N_q=(N_q)_{max}$ only.  These ratios are commented in the text as well as those 
in the tables
below.}
\begin{tabular}{|c|c|c|c|c|c|}
\hline
$(N_q)_{max}$ &$L$&$t_{\scriptscriptstyle\Sigma}$, 
$[s]$&$t'/t_{\scriptscriptstyle\Sigma}$
\\
\hline
21&11&$1.7\cdot10^{-3}$&68\\
\hline
22&8&$3.3\cdot10^{-3}$&110\\
\hline
23&9&$4.0\cdot10^{-3}$&120\\
\hline
24&8&$5.9\cdot10^{-3}$&130\\
\hline
25&9&$7.3\cdot10^{-3}$&150\\
\hline
26&10&$8.5\cdot10^{-3}$&170\\
\hline
27&9&$1.2\cdot10^{-2}$&210\\
\hline
28&10&$1.4\cdot10^{-2}$&240\\
\hline
\end{tabular}
\end{table}

 Comparison with the
running times $t'$ of Ref.~\cite{lit2} is presented as well
in the table. These times refer to the calculation at $N_q=(N_q)_{max}$ only.
Considering the listed ratios $t'/t_{\scriptscriptstyle\Sigma}$ 
one
should take into account that in the present calculation  a single
processor having the nominal performance less than 11~Gflops has been used.
 Whereas the calculation
in  Ref.~\cite{lit2} has been performed with a supercomputer and used in parallel
32 processors having the nominal performance about 16~Gflops each. 
Also the numbers of  corresponding brackets in our $N_q\le(N_q)_{max}$ computations
 exceed more than
twice the numbers of brackets in  the $N_q=(N_q)_{max}$ computations of Ref.~\cite{lit2}.

\begin{table}[ht]
\caption{The running times
for the
calculation of the sets of all existing brackets with even $N_q$ values
such that $N_q\le (N_q)_{max}$. 
The net numbers of such brackets 
are denoted as~$\cal N$. The running times in seconds for the computation
with \mbox{ALLOSBRAC} and \mbox{OSBRAC} are denoted as $t_1$ and $t_2$, 
respectively. The running time
 of Ref.~\cite{lit3} is denoted as $t'$.}
\begin{tabular}{|c|c|c|c|c|c|}
\hline
$(N_q)_{max}$&$\cal N$&$t_1$, $[s]$&$t_2$, $[s]$&$t_1/{\cal N}$, $[s]$&$t'/t_1$
\\
\hline
12&41424&$8.2\cdot10^{-4}$&$9.8\cdot10^{-4}$&$2.0\cdot10^{-8}$&33\\
\hline
16&243705&$4.1\cdot10^{-3}$&$4.5\cdot10^{-3}$&$1.7\cdot10^{-8}$&94\\
\hline
20&1040468&$1.6\cdot10^{-2}$&$1.7\cdot10^{-2}$&$1.5\cdot10^{-8}$&190\\
\hline
24&3555825&$5.3\cdot10^{-2}$&$5.3\cdot10^{-2}$&$1.5\cdot10^{-8}$&390\\
\hline
26&6165680&$9.4\cdot10^{-2}$&$9.0\cdot10^{-2}$&$1.5\cdot10^{-8}$&540\\
\hline
30&16743504&0.25&0.24&$1.5\cdot10^{-8}$&-\\
\hline
40&131288025&1.9&1.8&$1.5\cdot10^{-8}$&-\\
\hline
50&674606556&9.9&9.3&$1.5\cdot10^{-8}$&-\\
\hline
80&23084445209&-&330&-&-\\
\hline
\end{tabular}
\end{table}

In Table III the running times  for calculating all the existing brackets  with 
even $N_q$ values that do not exceed $(N_q)_{max}$ are listed. 
(These are brackets  
with all possible $L$ values.)
 The net numbers of such brackets 
 at given $(N_q)_{max}$  
are denoted as $\cal N$.  The running times, $t_1$ and $t_2$, refer, respectively, to use of
\mbox{ALLOSBRAC} and \mbox{OSBRAC}. At $(N_q)_{max}=80$ only the calculation
with \mbox{OSBRAC} is done because of insufficient memory. The running times $t_2$
are a bit larger at low  $(N_q)_{max}$. At higher $(N_q)_{max}$ they are somewhat smaller.
This is probably because of memory restrictions in the \mbox{ALLOSBRAC} case.
The running times per bracket prove to vary
slowly and they stabilize when $(N_q)_{max}$ increases. In the \mbox{OSBRAC} case
they equal $1.4\cdot 10^{-8}$ $s$ at $(N_q)_{max}\ge30$.

In the last column
the corresponding running times of Ref.~\cite{lit3}, 
deduced from the table there, are compared with the present running times.
In contrast  to our case, in the case of Ref.~\cite{lit3} the running times per bracket  sharply
increase with~$(N_q)_{max}$. Comparing the running times one should take into account
 that,  contrary to the present  single-processor
calculation with the computer described above, the calculation in Ref.~\cite{lit3} 
was performed in parallel using 12 processors. The same refers 
to the results in Table IV below.

\begin{table}[ht]
\caption{Accuracies   at calculating the sums (\re{del1}) and
(\re{del2}) and related running times in comparison with Ref.~\cite{lit3}.
Second column: the computed
values of the quantity $\delta_1(N_q)$ of Eq.~(\re{del2}). Third  column: the computed
values of the quantity $\delta_2(N_q)$ of Eq.~(\re{del1}). 
Fourth column: the quantity  of Eq.~(\re{del2}) or (\re{del1})
computed in Ref.~\cite{lit3} which is 
denoted as $\delta'(N_q)$. Fifth column: the running times 
$t_{\scriptscriptstyle\Sigma}$ for the calculation of
the sums $\sum_{N_q'}\delta_{1,2}(N_q')$. The summations proceed over 
$N'_q$ values of the same parity as $N_q$ and such that $N'_q\le N_q$.
Last column: the ratios of such running times $t'_{\scriptscriptstyle\Sigma}$ of
Ref.~\cite{lit3} to the above running times.}
\begin{tabular}{|c|c|c|c|c|c|}
\hline
$N_q$&$\delta_1(N_q)$&$\delta_2(N_q)$&
$\delta'(N_q)$ \cite{lit3}&$t_{\scriptscriptstyle\Sigma}$, $[s]$
&$t'_{\scriptscriptstyle\Sigma}/t_{\scriptscriptstyle\Sigma}$
\\
\hline
12&$2.55\cdot10^{-12}$&$1.3\cdot10^{-12}$&$3.2\cdot10^{-9}$
&$8.1\cdot10^{-4}$&$4.35\cdot10^{3}$\\
\hline
16&$1.1\cdot10^{-11}$&$4.6\cdot10^{-12}$&$1.5\cdot10^{-6}$&$4.0\cdot10^{-3}$
&$2.2\cdot10^{4}$\\
\hline
20&$3.8\cdot10^{-11}$&$1.2\cdot10^{-11}$&$6.1\cdot10^{-4}$&$1.6\cdot10^{-2}$
&$8.8\cdot10^{4}$\\
\hline
25&$1.65\cdot10^{-10}$&$3.2\cdot10^{-11}$&0.72&
$7.0\cdot10^{-2}$&$3.6\cdot10^{5}$\\
\hline
30&$7.1\cdot10^{-10}$&$6.7\cdot10^{-11}$&-&0.25&-\\
\hline
35&$3.5\cdot10^{-9}$&$1.2\cdot10^{-10}$&-&0.75&-\\
\hline
\end{tabular}
\end{table}
 
In Table IV the  quantities equal to zero
\be
\delta_1(N_q)=\sum_{i,i',L;N_q=const}\bigl|\bigl[\sum_j
\langle j|i\rangle_L^\varphi\langle j|i'\rangle_L^\varphi\bigr]
-\delta_{ii'}\bigr|\la{del2}
\ee
\be
\delta_2(N_q)=\bigl|\sum_{i,i',L;N_q=const}\bigl[\sum_j
\langle j|i\rangle_L^\varphi\langle j|i'\rangle_L^\varphi\bigr]
-\delta_{ii'}\bigr|\la{del1}
\ee
are calculated. Here the summations  proceed at given  $N_q$ values and the other notation 
is as in Eq.~(\re{del}). The computation was done with \mbox{ALLOSBRAC}. 
In the fourth column the
values  of the sum obtained in Ref.~\cite{lit3} are listed. They are denoted 
as $\delta'$.  It is not known whether this was
the sum (\re{del2}) or (\re{del1}). The calculation
completely loses stability at \mbox{$N_q>20$}. 
In the fifth column 
the running times are given for the calculation of the sums $\sum_{N_q'}\delta_{1,2}(N_q')$.
The summations proceed over
$N'_q$ values of the same parity as $N_q$ and such that $N'_q\le N_q$.
These running times are denoted as $t_{\scriptscriptstyle\Sigma}$.
In the last column the ratios of such running times~$t'_{\scriptscriptstyle\Sigma}$ of
Ref.~\cite{lit3}, deduced from the table there, to the above running times
are presented.
As in the tables above, these ratios strongly increase as $N_q$ increases.

\begin{table}[ht]
\caption{Dependence of the calculated quantity (\re{eps}) on $N_q$.}
\begin{tabular}{|c|c|}
\hline
$N_q$&$\delta_{rel}$\\
\hline
10&$1.6\cdot10^{-15}$\\
\hline
20&$5.4\cdot10^{-15}$\\
\hline
30&$1.6\cdot10^{-13}$\\
\hline
40&$4.0\cdot10^{-12}$\\
\hline
50&$1.1\cdot10^{-10}$\\
\hline
60&$3.3\cdot10^{-9}$\\
\hline
70&$1.1\cdot10^{-7}$\\
\hline
80&$2.7\cdot10^{-6}$\\
\hline
\end{tabular}
\end{table}

The quantity equal to zero
\be \delta_{rel}={\textstyle \frac{1}{2}}
\max_{i,L}\bigl|\bigl[\sum_{\,j;N_q=const}\bigl(\langle j|i\rangle_L^\varphi\bigr)^2
\bigr]-1\bigr|
\la{eps}\ee 
is also calculated. The notation is as above. This quantity represents an estimate of the relative
error of computed brackets  at a given $N_q$ value. Indeed, 
in the presence  of errors $\Delta(j,i,L)$  in  values of the brackets one has  
\[ {\textstyle \frac{1}{2}}\bigl\{\bigl[\sum_{\,j;N_q=const}
\bigl(\langle j|i\rangle_L^\varphi\bigr)^2
\bigr]-1\bigr\}\simeq\sum_{\,j;N_q=const}\bigl(\langle j|i\rangle_L^\varphi\bigr)^2
\frac{\Delta(j,i,L)}{\langle j|i\rangle_L^\varphi}.\]
The latter quantity is the relative error averaged over the 
$(\langle j|i\rangle_L^\varphi)^2$ distrbution.   
The obtained values of the quantity (\re{eps}) are presented in Table V. The results at
the highest
$N_q$ values indicate the limitations of the double precision computation.  

In conclusion, the programs to calculate oscillator brackets have been created.
The listed runs show that the programs are fast and 
results they produce are accurate up to very high
oscillator excitations. Contrary to the programs based on the explicit expression
for the brackets, 
the  amount of computations per
bracket  practically does not change as quantum numbers increase. 
Output arrays of the brackets  are quite convenient for majority
of applications. 
These arrays are made compact due to use  of the
suitable combinations of partial orbital momenta as array arguments.
The programs are easy to implement and follow.

\end{document}